# Self- and Mutual Inductance of NbN and Bilayer NbN/Nb Inductors in Planarized Fabrication Process With Nb Ground Planes

Sergey K. Tolpygo, *Senior Member, IEEE,* Evan B. Golden, Terence J. Weir, Vladimir Bolkhovsky, and Ravi Rastogi

*Abstract*—We present measurements of the self- and mutual inductance of NbN and bilayer NbN/Nb inductors (microstrips, striplines, serpentines, etc.) with Nb ground plane(s) fabricated in an advanced process for superconductor electronics developed at MIT Lincoln Laboratory. In this process, the signal traces of logic cell inductors are made either of a 200-nm NbN layer with $T_c$ ≈15 K or of an in-situ deposited NbN/Nb bilayer, replacing a 200-nm Nb layer M6 in the standard SFQ5ee process with nine superconducting layers on 200-mm wafers. Nb ground planes were preserved to maintain a high level of interlayer shielding and low intralayer mutual coupling. A two-step patterning of the top Nb and the bottom NbN layers of the NbN/Nb bilayer allows to create inductors in a very wide range of linear inductance values, from low values ~ 0.4 pH/μm typical for Nb geometrical inductors to ~ 35 pH/μm typical to thin-film kinetic inductors. Mutual inductance of NbN and Nb inductors, of NbN inductors, and of bilayer inductors is the same as between two Nb inductors with the same geometry and placement between the ground planes, *i.e.*, mutual inductance does not depend on superconducting properties of the signal traces in the studied range of linewidths. We measured magnetic field penetration depth and kinetic inductance of NbN films with thickness $t$ =200 nm to be $\lambda$ = 491±5 nm and 1.51 pH/sq, and 2.06 pH/sq at $t$ =150 nm. The kinetic inductance was found to be larger than that expected for superconductors with short mean free path, indicating a reduction in the superfluid density in reactively sputtered NbN films, likely due to disorder and carrier localization effects. Kinetic inductance associated with right-angled bends of the NbN inductors was found to be negligible at linewidths $w < \lambda^2/t$, indicating a very small current crowding in structures with superconducting ground plane(s). Implementation of NbN and NbN/Nb bilayer inductors allows for a significant increase in the circuit density and integration scale of superconductor digital electronics.

## I. Introduction

ADVANTAGES of superconductor digital electronics over semiconductor electronics in clock frequency and energy efficiency are well known. However, a three to four orders of magnitude lower scale of integration of superconductor electronics (SCE) compared to CMOS impedes SCE applications. The main limiting factor to the integration scale is the size of Josephson junctions (JJs) and inductors in logic and memory cells [1].

The current trend of SCE development is towards lowering energy dissipation, which is achieved by decreasing the typical critical current $I_c$ of JJs and hence of their area. This in turn increases inductance values, and area, of all inductors in the cells because their typical inductance $L$ relates to $I_c$ through the cell design parameter $\beta_L = 2\pi L I_c/\Phi_0$ determined in the cell simulation and optimization [2]. Another trend of the last decade is proliferation of multiphase ac clocked logics utilizing either bipolar switching of JJs as in RQL logic [3] or quasi-adiabatic changes in the flux state of logic cells as in AQFP [4], [5]. In addition to the regular cell inductors, ac-clocked circuits require efficient cell transformers (mutual) inductors to couple in the ac excitation. These transformers are poorly scalable as explained in [6], [7].

It is also well known [1] that area occupied by cell inductors can be substantially decreased if Nb inductors in integrated circuits, which are mainly magnetic (geometrical) inductors, could be replaced by kinetic inductors – thin films of materials with large magnetic field penetration depth $\lambda \gg t$, where $t$ is the film thickness, giving rise to kinetic inductance $L_{K,sq} = \mu_0 \lambda^2/t \approx 1.257 \lambda^2/t$ (in pH per square for $\lambda$ and $t$ in μm) that is much larger than magnetic inductance $L_m$. Such materials are also well known, e.g., thin films of NbN, NbTiN, Mo$_2$N [2], and can theoretically replace Nb layers in a multi-layer stack of the existing fabrication processes [8]–[10].

However, implementing kinetic inductors would significantly decrease mutual inductance in transformers because reducing the length of inductors would also decrease their mutual running length. So, superconductor logics and memories utilizing ac excitation may not necessarily benefit from implementation of the kinetic inductors.

Manuscript receipt and acceptance dates will be inserted here. This material is based upon work supported by the Under Secretary of Defense for Research and Engineering under Air Force Contract No. FA8702-15-D-0001. *(Corresponding author: Sergey K. Tolpygo. E-mail: sergey.tolpygo@ll.mit.edu.)*

All authors are with Lincoln Laboratory, Massachusetts Institute of Technology, Lexington, MA 02421, USA (e-mails: evan.golden@ll.mit.edu, weir@ll.mit.edu, blokv@ll.mit.edu, ravi.rastogi@ll.mit.edu ).

Color versions of one or more of the figures in this paper are available online at http://ieeexplore.ieee.org.
Digital Object Identifier will be inserted here upon acceptance.

It is also clear that all Nb layers in the process stack cannot be replaced by superconducting materials having much larger penetration depth than $\lambda_{Nb}$= 90 nm because this would dramatically compromise screening properties of the ground planes, increase parasitic coupling between all inductors, increase impedance of and delay in passive transmission lines (PTLs), etc. Therefore, materials with high kinetic inductance can replace Nb only on some dedicated inductor layers whereas Nb should be preserved on all layers used for ground planes and PTLs.

Despite a clear need for kinetic inductors for increasing integration scale, we are not aware of their implementations in logic cells of superconductor integrated circuits. In the SFQ5ee process at MIT LL, 40-nm-thick films of $Mo_2N$ with inductance of 8 pH/sq [11] have been used only as rf choke inductors in a voltage biasing scheme of ERSFQ circuits [12], [13], [14].

We describe properties of inductors made of 200-nm-thick NbN films and bilayers NbN/Nb incorporated as the dedicated inductor layer in the layer stack of the SFQ5ee process [9], [10] with 8 planarized niobium layers, replacing Nb layer M6.

The use of NbN/Nb bilayers allows us to combine kinetic inductors and magnetic (geometrical) inductors on the same circuit level and thereby preserve the required mutual inductance in transformers and provide compact kinetic inductors in logic cells.

## II. Fabrication Process and Measurements

### A. The SFQ5ee Process with NbN Layer M6

Process flow of the standard SFQ5ee process with 8 niobium planarized layers on 200-mm wafers, described in [9], [10], was used up to the metal layer M6, the layer interconnecting Josephson junctions and shunt resistors; see [15, Fig. 1], [9, Fig. 1], [10, Fig. 1]. At this step, a 200-nm-thick NbN layer was deposited (instead of Nb) at 200 °C, using reactive sputtering of Nb in $Ar/N_2$ flow, in one of the deposition chambers of an Endura PVD cluster tool. After the standard 248-nm photolithography, NbN layer M6 was patterned using high-density plasma etching in $Cl_2$-based chemistry. The patterned layer was planarized using $SiO_2$ deposition and chemical mechanical polishing, forming interlayer dielectric, I6. Two dielectric thicknesses were used: 200 nm on wafer #9 (w9), as in the standard process; and 100 nm on wafer #10 (w10). The latter was done to increase mutual coupling between inductors on the layers M6 and M7. Then, Nb layer M7 was deposited and patterned as in the standard SFQ5ee process, followed by the wafer surface passivation and contact pad metallization steps completing the process flow. This process modification does not require any additional photomasks or processing steps in comparison to the SFQ5ee process.

### B. The SFQ5ee Process with NbN/Nb Bilayer on Layer M6

Similar to II.A, after reaching deposition step of the layer M6, it was deposited in-situ as NbN/Nb bilayer with total thickness of 200 nm. Thicknesses of the individual layers $t_{NbN}$ and $t_{Nb}$ were varied to investigate various combination, e.g., $t_{NbN}$ =150 nm, $t_{Nb}$ = 100 nm, and $t_{NbN} = t_{Nb}$ =100 nm.

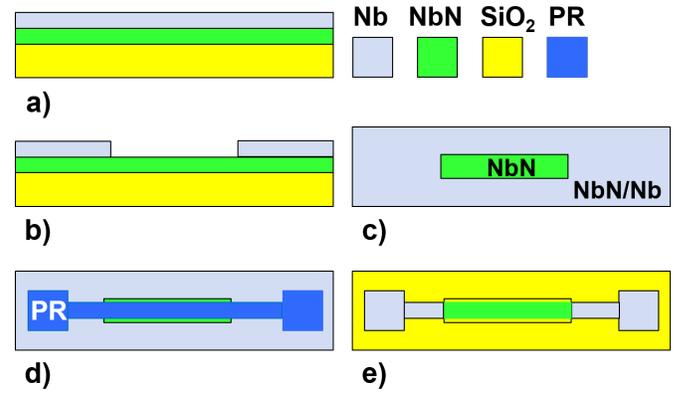

Fig. 1. Processing of the NbN/Nb bilayer: a) cross section of a bilayer deposited on a planarized layer of $SiO_2$ interlayer dielectric, I5 (between layers M5 and M6); b) patterning of the top Nb layer, using a dark field mask M6a and selective etching of Nb over NbN; c) top view of the etched structure; d) second photolithography step uses a clear field mask M6 to define the inductor shape; e) top view after etching NbN/Nb bilayer (not protected by the photoresist mask) down to the $SiO_2$ dielectric. These steps produce a composite inductor consisting of NbN central part (green), aligned in the window etched in Nb in the step c), and the NbN/Nb bilayer parts (grey) providing low inductance connections to the high kinetic inductance NbN. The NbN and NbN/Nb parts of the composite inductors may have the same or different widths, depending on the circuit design requirements. Wider area of the bilayer shown at the ends of the inductor in d) and e) are used for connections to JJs and interlayer vias.

Patterning of the bilayer is shown in Fig. 1 and done in two photolithography and two etching steps to pattern individually the top Nb and the bottom NbN layers of the bilayer. The first photolithography step uses a dark field mask M6a and a positive tone photoresist (PR). This step does not exist in the standard SFQ5ee process. It creates unmasked etch windows for etching the top Nb of the bilayer in the regions where NbN inductors are going to be placed. Selective etching of the top Nb in the unmasked areas is done using an optical emission end-point detector and a 2- to 3-nm-thick etch-stop (ES) layer deposited on the surface of NbN. That is the bilayer was actually deposited as a trilayer, e.g., NbN(150)/ES(3)/Nb(100). There are several metals which can be used as a hard ES layer, e.g., Ti, Al, Pt, Cu, Au, etc., depending on the etch chemistry used. Then, the second photolithography is done using a clear field M6 mask and the same positive PR; see Fig.1d. Finally, the bilayer is etched off from the entire unmasked area in one etching step, stopping on the underlaying $SiO_2$; see Fig. 1e. As a result of this two-step patterning, we formed the variable thickness, composite inductors consisting of a thin NbN part or several parts and thick parts composed of the full bilayer. After the bilayer processing, the process layer stack is completed as in II.A by depositing and planarizing the interlayer SiO₂ dielectric I6, and depositing and patterning the last Nb layer M7.

TABLE I
PARAMETERS OF LAYERS IN MEASURED INDUCTORS

| Parameter | w9 (nm) | w10 (nm) |
|---|---|---|
| Nb ground plane layer M4 thickness (nm) | 200 | 200 |
| Nb layer M5 thickness | 135 | 135 |
| Dielectric thickness between layers M4 and M5 | 200 | 200 |
| Dielectric thickness between layers M4 and M6 | 615 | 615 |
| NbN layer M6 thickness | 200 | 200 |
| Dielectric thickness between layers M6 and M7 | 200 | 100 |
| Dielectric thickness between layers M4 and M7 | 1015 | 915 |
| Nb layer M7 thickness | 200 | 200 |
| Magnetic field penetration depth in Nb films | 90 | 90 |

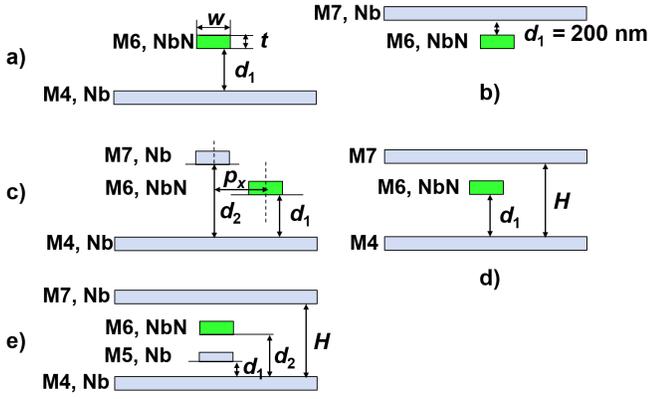

Fig. 2. Cross sections of the inductors studied: a) microstrips M6aM4 (standing for a microstrip with signal trace on the layer M6 above the M4 ground plane); b) inverted microstrip M6bM7 ("bM7" stands for below the M7 ground plane); c) mutual coupling of microstrip inductors M6aM4 and M7aM4 , $p_x$ is the distance between geometrical centers of the signal traces in the horizontal, $x$-direction; d) stripline inductor M6aM4bM7; e) mutual coupling of striplines M5aM4bM7 and M6aM4bM7; distance between their geometrical center was varied. Dielectric thicknesses entering analytical expressions in the text are also indicated: $d_1$ and $d_2$ are the dielectric thicknesses whose values depend on the type of the inductors and given in Table I.

### C. Inductance and Mutual Inductance Measurements

Inductance and mutual measurements were done using a SQUID-based method [16] and the integrated circuit developed in [17, Fig. 4], and discussed in detail in [6], [18]. Extensive data on Nb microstrip and stripline inductors as well as analytical expressions were given in [6]. In all cases, we measured the differential inductance of two arms of the SQUID, differing only by the length of inductors, in order to remove all parasitic contributions. The measured values were normalized to the length difference, 45 μm, giving self-inductance, $L_l$ or mutual inductance, $M_l$ per unit length. The nominal thicknesses of dielectric and superconducting layers involved in the studied inductors are given in Table I. Cross sections of the typical inductors and mutual inductors studied are sketched in Fig. 2.

## III. SELF- AND MUTUAL INDUCTANCE OF NbN INDUCTORS

### A. Inductance of NbN Microstrips with Nb Ground Planes

Inductance of NbN microstrips M6aM4, Fig. 2a, and M6bM7, Fig. 2b, per unit length is shown in Fig. 3 for microstrips of various widths, $w$. For comparison, we give data for Nb microstrips of the same types, fabricated on a different wafer in the same process run. Obviously, inductance of the microstrips with NbN signal traces is much larger than of microstrips with the same geometry but with Nb traces.

According to [6], inductance of superconducting microstrips per unit lengths is given by

$$L_l = \frac{\mu\mu_0}{4\pi}\ln\left[1 + \frac{4\left(d_1+\frac{t_1}{2}+\lambda\right)^2}{0.2235^2(w+t_1)^2}\right] + \frac{\mu_0\lambda_1^2}{t_1 w}, \quad (1)$$

where $d_1$ is the dielectric thickness between the signal strip and the ground plane, $\lambda$ and $\lambda_1$ are magnetic field penetration depths in, respectively, the ground plane(s) and the signal trace materials; $\mu_0 = 4\pi \times 10^{-7}$ H·m$^{-1}$ and $\mu$ are, respectively, the vacuum and relative magnetic permeability; the latter is assumed equal unity in all the calculations hereafter. The first term in (1) is magnetic part of the inductance, which depends on the penetration depth in the ground plane. The second term is the kinetic inductance for a uniform current distribution in the signal trace, which is the case since $\lambda_1 \gg t$, and $\lambda_1^2/t \gg w$, and also because the superconducting ground plane(s) promote the uniform current distribution.

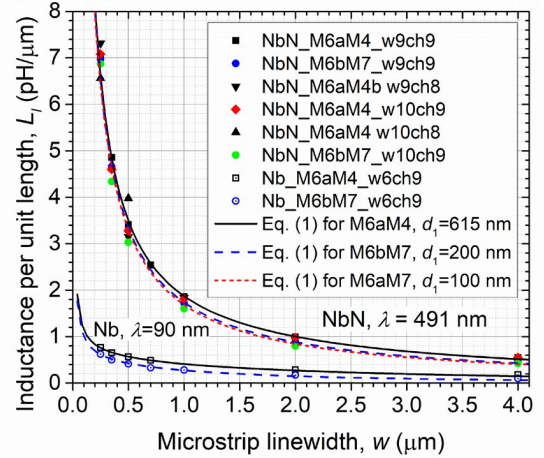

Fig. 3. Inductance per unit length of microstrips M6aM4 and inverted microstrips M6bM7 using 200-nm NbN layer M6 and Nb ground planes M4 or M7 as a function of linewidth of the NbN signal traces, $w$. Data for two wafers, w9 and w10, and two 5 mm ×5 mm chips, ch8 and ch9, at two locations on 200-mm wafers are given. For comparison, the data are also given for all-Nb microstrips co-fabricated in the same process run on wafer #6 (w6) on the same chips and locations on the wafer. Solid and dash curves are fits to (1), giving magnetic field penetration depth in NbN $\lambda_1$ =491±5 nm at magnetic field penetration depth in Nb ground planes of $\lambda$ =90 nm; see the two bottom curves for Nb. Short dash magenta curve is also a fit to (1) but at a different dielectric thickness $d_1$ =100 nm between the layers M6 and M7 on w10.

Fitting to (1) of the experimental data on the microstrips with three different values of $d_1$ =615 nm, 200 nm, and 100 nm is shown in Fig. 3, and gives a mean value of magnetic field penetration depth in NbN films $\lambda_1$ =491±5 nm. This translates into a sheet kinetic inductance of 200-nm NbN films $L_{K,sq}$ =1.51 pH/sq. Consequently, 150-nm and 100-nm-thick NbN films should have sheet kinetic inductances of 2.01 and 3.02 pH/sq, respectively. For comparison, a 200-nm-thick Nb film has kinetic inductance of 0.051 pH/sq.

### B. Inductance of NbN Striplines M6aM4bM7 with Nb Ground Planes

Inductance per unit length of NbN striplines M6aM4bM7 with Nb ground planes M4 and M7 is shown in Fig. 4, along with fully Nb striplines for comparison. Similar to the microstrips in Fig. 3, inductance of NbN striplines is many times larger than of the Nb striplines with the same dimensions.

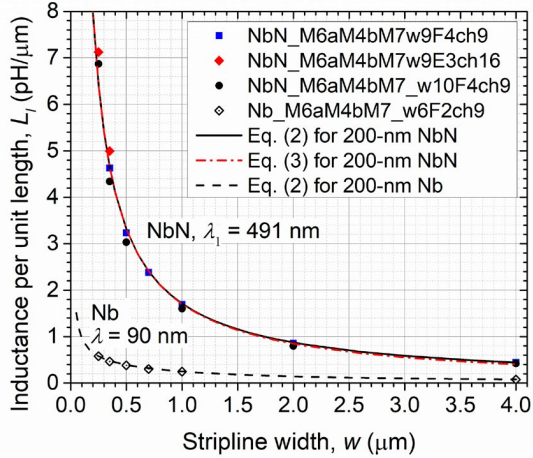

Fig. 4. Inductance per unit length of striplines M6aM4bM7 (Fig. 2d) using 200-nm NbN layer M6 and Nb ground planes M4 and M7 as a function of linewidth of the NbN signal traces, $w$. Data on two wafers, w9 and w10, and two 5 mm ×5 mm chips, ch9 and ch16, at two locations, F4 and E3, on 200-mm wafers are given. For comparison, the data are also given for Nb striplines co-fabricated in the same process run on wafer #6 (w6). Solid black curve is (2) at magnetic field penetration depth in NbN $\lambda_1$ =491 nm and $\lambda$ =90 nm in all the Nb layers; $H$ = 1015 nm, $d_1$ =615 nm; see Table I. Red dash-dot curve is (3) at the same parameters. Dash curve is (2) for Nb striplines at $\lambda_1 = \lambda$ =90 nm.

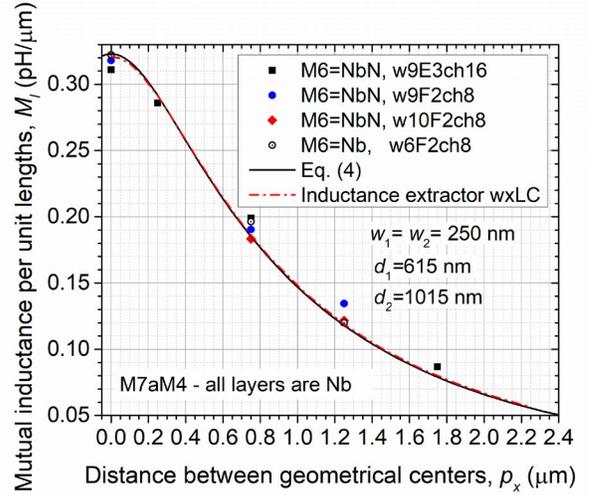

Fig. 5. Mutual inductance per unit length of Nb microstrips M7aM4 and NbN microstrips M6aM4 with equal width $w$ =250 nm as a function of horizontal spacing between the geometrical centers of microstrip cross sections, $p_x$; see Fig. 2c. For comparison, the data are also given for all-Nb microstrips co-fabricated in the same process run on wafer #6 (w6). Solid black curve is (4) at magnetic field penetration depth in the Nb ground plane $\lambda$ =90 nm and $d_1$ =615 nm for the M6 layer, and $d_2$ =1015 nm for the M7 layer; see Table I. Red dash-dot curve shows numerical simulation done using inductance extractor wxLC [24] at the same parameters.

Superconducting stripline inductance is given in [6] by

$$L_l = \frac{\mu\mu_0}{4\pi}\ln\left(1+\frac{\sin^2\frac{\pi\left(d_1+\frac{t_1}{2}+\lambda\right)}{H+2\lambda}}{\sinh^2\frac{\pi r_{eq}}{2(H+2\lambda)}}\right)+\frac{\mu_0}{8\pi}+\frac{\mu_0\lambda_1^2}{t_1 w}, \qquad (2)$$

where $r_{eq}$ is the equivalent radius of the rectangular cross section given in [6, (23a)] and [6, (23b)], and $H$ is the dielectric thickness between the ground planes; see Fig. 2d. The sum of the first two terms in (2) is magnetic inductance of the stripline, and the third term is kinetic inductance of the signal strip. This expression with the penetration depths determined from the fits in Fig. 3, is plotted in Fig. 4 for NbN striplines (solid black curve) and Nb striplines (dash black curve), and gives an excellent description of the data.

Instead of the equivalent radius approximation (2), magnetic component of the stripline inductance can be also expressed using Maxwell's geometric mean distance method [19]–[21], calculating self-inductance of a wire with rectangular cross section as mutual inductance of the wire with itself placed at the geometrical mean distance, giving,

$$L_l = \frac{\mu\mu_0}{4\pi}\ln\left(1+\frac{\sin^2\frac{\pi\left(d_1+\frac{t_1}{2}+\lambda\right)}{H+2\lambda}}{\sinh^2\frac{0.2235\pi(w+t)}{2(H+2\lambda)}}\right)+\frac{\mu_0\lambda_1^2}{t_1 w}, \qquad (3)$$

where $0.2235(w+t_1)$ is Maxwell's geometrical mean distance for a rectangular cross section; see [20, (124)]. Expression (3) is also plotted in Fig. 4 by the dash-dot red curve. It gives nearly identical values to (2) up to $w$ =1 μm and somewhat lower inductance at larger linewidth; the difference is −3% at $w$ = 2 μm, gradually increasing to −9% at $w$ = 4 μm. Within the accuracy of the experimental data and parameters spread across 200-mm wafers, both (2) and (3) can be used to calculate inductance of NbN striplines in the range of linewidths important for inductors in superconductor integrated circuits, $w \leq 2$ μm.

### C. Mutual Inductance of NbN and Nb Microstrips

Mutual inductance between Nb microstrip M7aM4 and NbN microstrip M6aM4 is shown in Fig. 5 as a function of horizontal distance between the geometrical centers of the microstrips' cross sections, $p_x$; see Fig. 2c. The linewidths of both microstrips are equal and $w_1 = w_2$ =250 nm. For comparison, also shown is mutual inductance for all-Nb microstrips of the same width (open dots) fabricated on the wafer #6. Within error of the measurements and wafer-to-wafer variation, there is no difference between the mutual inductance of the NbN−Nb pair and Nb–Nb pair of the microstrips. The measurements confirm that the mutual inductance matrix is symmetrical in all the cases described hereafter.

Mutual inductance of two superconducting microstrips with thicknesses $t_1$ and $t_2$, dielectric thicknesses $d_1$ and $d_2$, and linewidths $w_1, w_2 \lesssim |d_2 - d_1|$ is given in [6] by

$$M_l = \frac{\mu\mu_0}{4\pi}\ln\left[1+\frac{4\left(d_1+\lambda+\frac{t_1}{2}\right)\cdot\left(d_2+\lambda+\frac{t_2}{2}\right)}{p_x^2+(d_2-d_1+\frac{t_2-t_1}{2})^2}\right]. \qquad (4)$$

It is plotted in Fig. 5 by the solid black curve and describes the experimental data very well. For comparison, we also show by the dash-dot red curve the results of mutual inductance extraction using wxLC software for a transmission line approximation developed by M. Khapaev [24]. At small linewidths, (4) and the numerical results are virtually indistinguishable. We note that $M_l$ decays slowly with increasing the distance between the microstrips, only as a second power of the distance. This gives rise to a large parasitic mutual coupling of inductors in dense circuits.

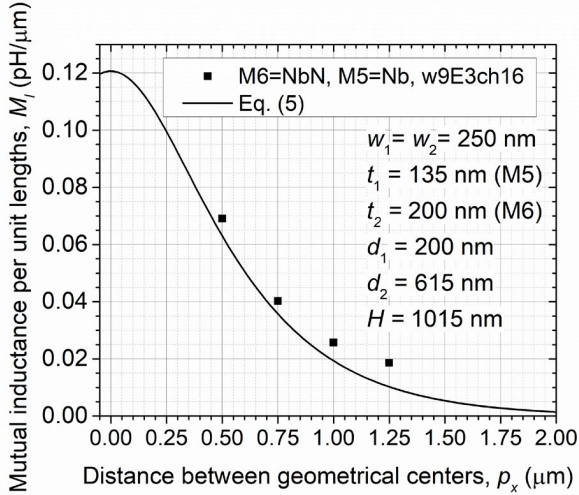

Fig. 6. Mutual inductance per unit length of Nb stripline M5aM4bM7 and NbN stripline M6aM4bM7 with equal width $w$ =250 nm as a function of horizontal spacing between the geometrical centers of their cross sections, $p_x$; see Fig. 2e. Solid black curve is (5) at magnetic field penetration depth in the Nb ground planes $\lambda$ =90 nm and $d_1$ =200 nm, $d_2$ =615 nm, and $H$ =1015 nm; see Table I.

### D. Mutual Inductance of NbN and Nb Striplines

Mutual inductance of NbN striplines M6aM4bM7 and Nb striplines M5aM4bM, see Fig. 2e, of equal width $w_1 = w_2$ =250 nm is shown in Fig. 6. The measurements confirm that the mutual inductance matrix is symmetrical. Note that mutual inductance of the striplines is much smaller than mutual inductance of the microstrips in Fig. 5 and is more difficult to measure as a result.

Mutual inductance of two superconducting striplines is given in [6] by

$$M_l = \frac{\mu\mu_0}{4\pi} \ln \frac{\cosh\frac{\pi p_x}{H+2\lambda} - \cos\frac{\pi(d_1+d_2+t_1+t_2+2\lambda)}{H+2\lambda}}{\cosh\frac{\pi p_x}{H+2\lambda} - \cos\frac{\pi(d_2-d_1+t_2-t_1)}{H+2\lambda}}. \quad (5)$$

It is plotted in Fig. 6 for the parameters of the striplines used and provides a reasonable description of the experimental data. We believe that progressively increasing difference between the experiment and (5) with increasing $p_x$ is likely a result of experimental difficulties with measuring small mutual inductances. This requires applying large modulation currents through the primary inductor M5 forming a transformer with a large NbN inductor M6 in the SQUID arm.

### E. Kinetic Inductance of 90-degree Bends of NbN Inductors

The method of measuring inductance of bends of the thin-film inductors was described in [6]. It is based on measuring the difference of two inductors, $\Delta L_{meas}$, with the same total length but vastly different number of bends. The inductance associated with a corner square of a 90-degree bend is given by

$$L_c = L_{sq} - |\Delta L_{meas}/\Delta N|, \quad (6)$$

if the mutual inductance between the straight sections of the bended wire can neglected; here $L_{sq}$ is the sheet inductance per square and $\Delta N$ is the difference in the number of bends in the two inductors; see [6, eq. (53)].

Inductance is a complete analog of electrical resistance. If current distribution in the thin-film inductor is the same as in the thin-film resistor, the ratio $\alpha = L_c/L_{sq}$ should the same as the ratio of the bend resistance to the sheet resistance, $\alpha = R_{bend}/R_{sh}$. The latter depends on the bend shape, i.e., the amount of current crowding near the inner part of the bend, and for many shapes was calculated in [22], [23]. For a 90-degree bend, the corner square contribution is $\alpha = 0.56$ [23]. Hence, if the mutual inductance between the straight sections of the meander can be neglected, the kinetic inductance of a strip with $N$ bends is expected to be $L_K = \frac{\mu_0 \lambda_1^2}{t}[\frac{l}{w} - (1-\alpha)N]$ with $\alpha$ =0.56 for the 90-degree bends.

We used a set of six pairs of NbN stripline meander inductors M6aM4bM7 with $\Delta N$ =40 and linewidths from 0.25 µm to 2.0 µm. Geometrical parameters of this set are identical to the first six inductors given in [6, Table II]. The layers M4 and M7 were Nb layers as in Secs. *A-D*, the NbN layer M6 was 200 nm thick.

For the inductance per unit length of NbN striplines with Nb ground planes the following relations hold: $L_l^{NbN} \gg L_l^{Nb} > L_{l,G} > M_l$, where $L_l^{Nb}$ is inductance per unit length of the Nb stripline with the same width $w$ as the NbN stripline, $L_{l,G}$ is the geometrical inductance of the striplines, and $M_l$ is the mutual inductance of the two equal-width striplines per unit length. Hence, the mutual inductance between the straight sections of the NbN meandered inductors can be neglected with respect to their self-inductance, and the inductance of the corner square of the right-angled bend can be calculated using (6); see [6].

The dependence of the corner inductance of the right-angled bends of NbN striplines M6aM4bM7 on the linewidth is shown in Fig. 7. Since inductance of NbN striplines is mainly kinetic, see Fig. 4, the extracted corner inductance is almost entirely the kinetic inductance associated with the bend. It is clear from Fig. 7, that the total (and the kinetic) inductance of the bend is negligible, $\alpha \ll 1$, at small linewidths $w \ll \lambda_1^2/t \approx 1.25$ µm, contrary to the expectations based on [23]; $\alpha$ asymptotically approaches the theoretical value $\alpha = 0.56$ only at $w \gg \lambda_1^2/t$.

For comparison, we present also the corner inductance of the co-fabricated Nb striplines, whose inductance is mainly magnetic, extracted using (6). Small negative value of $\alpha$ for the Nb striplines at the smallest linewidth is the result of neglecting the mutual inductance between the straight parallel sections of the meander. It needs to be accounted for at small linewidths and small spacings between the meander turns, as was explained in [6]. Using a more general formula [6, eq. (58)]

$$L_c = L_{sq} - |\frac{\Delta L_{meas}}{\Delta N}| + \frac{2M_l \Delta y(\frac{N}{2}-1)}{\Delta N} \quad (6a)$$

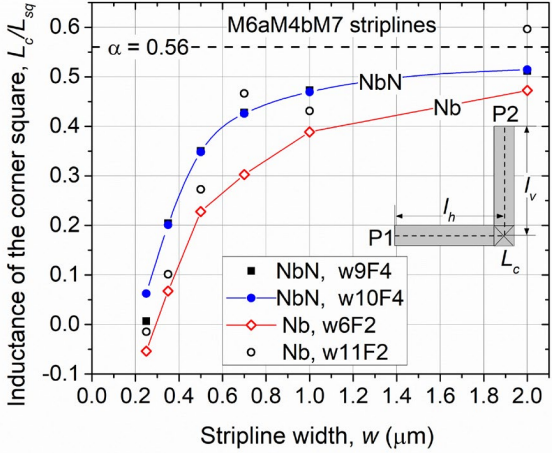

Fig. 7. Normalized inductance of the corner square, $L_c$, of 90-degree bends of meandered striplines M6aM4bM7 when the M6 strip is either NbN or Nb, while both M4 and M7 are Nb ground planes; point connections are to guide the eye. The geometrical dimensions of the NbN and Nb striplines are identical; they were given in [6, Table II]. Inset: a top view of the meander section with length $l = l_h + l_v$ measured along the central line of the bended strip. Inductance of the corner square is defined as $L_c = L_{P1,P2} - L_l(l_h + l_v - w)$, where $L_{P1,P2}$ is the inductance measured between ports P1 and P2 placed at the ends of the strip far away from the bend; $L_{sq} \equiv L_l w$.

changes the negative values of $L_c$ to small positive values, where $\Delta y = l_v - w$ is the mutual running length of the parallel vertical sections of the meander with $N$ corners.

Accounting for the mutual inductances inside the meander does not change the main result that, for both the kinetic and magnetic inductors, the 90-degree bend inductance is much smaller than the value $\alpha = 0.56$ following from the current crowding near the inner corner of the bend [23]. Most likely, the current distribution near the bends is much more uniform in the presence of the superconducting ground planes than that obtained in [22], [23] and in the later studies of the critical currents of bended wires, using conformal mapping of the single-layer wire bends; see, e.g., [58], [59]. To summarize, the current crowding effect on the inductance of bends and turns can be neglected for thin-film ($t < \lambda_1$) kinetic inductors with $w \ll \lambda_1^2/t$ and magnetic inductors with $w < d_1$ in the presence of superconducting ground plane(s).

## IV. Inductance of Composite Inductors NbN + NbN/Nb

In Table II and Table III, we present inductance data for the composite NbN + NbN/Nb microstrips consisting of a NbN/Nb bilayer with the top Nb layer etched away on length $l_{NbN}$, a part of the microstrip total length $l$, as described in Fig. 1. We investigated several types of composite inductors with several thicknesses of the NbN and Nb layers, $t_{NbN}$ and $t_{Nb}$, uniform width $w$, and various lengths $l_{NbN}$ and $l$.

In order to calculate inductance of the composite inductors, we used a procedure proposed in [7]. Briefly, we calculated inductance per unit length of the bilayers, $L_{l,bi}$, as a parallel combination of inductances of the bottom NbN layer, $L_{l,NbN}$, and the top Nb layer, $L_{l,Nb}$, per unit length

$$L_{l,bi} = \frac{L_{l,NbN} L_{l,Nb} - M_{l,NbN\text{-}Nb}^2}{L_{l,NbN} + L_{l,Nb} - 2M_{l,NbN\text{-}Nb}}, \quad (7)$$

where $M_{l,NbN\text{-}Nb}$ is the mutual inductance between the layers in the bilayer per unit length. For verification, we applied (7) to Nb/Nb bilayers and obtained the bilayer inductance within 0.2% from the expected value – the inductance of the single-layer Nb inductors with thickness equal to the bilayer total thickness.

Next, we calculated the total inductance of the composite inductors as a series connection of the bilayer inductance, $L_{bi}$, the NbN geometrical inductance, $L_G^{NbN}$, and the NbN kinetic inductance, $L_K^{NbN}$

$$L_{calc} = L_{bi} + L_G^{NbN} + L_K^{NbN} = (l - l_{NbN})L_{l,bi} + l_{NbN}L_{G,l}^{NbN} + \frac{l_{NbN}}{w}L_{K,sq}^{NbN}, \quad (8)$$

where $L_{G,l}^{NbN}$ is the geometrical inductance of the NbN layer per unit length, $L_{K,sq}^{NbN} = \mu_0 \lambda_{NbN}^2 / t_{NbN}$ is the NbN layer kinetic inductance per square, and $l_{NbN}/w = N_{sq}$ is the number of squares of NbN film with the etched top Nb layer. In (8), we neglected a potential inductance associated with electric current redistribution between the bilayer and the NbN layer near the boundaries of the etched Nb region.

Unfortunately, inductances of the individual layers in the bilayers and the penetration depths in them, $\lambda_{Nb}$ and $\lambda_{NbN}$, cannot be measured independently. Therefore, we used $\lambda_{Nb} = 90$ nm [6], [17], [18], the same value as for all Nb ground planes, $\lambda$, and treated $\lambda_{NbN}$ as the only adjustable parameter. It was adjusted to minimize the mean absolute difference (MAD) $\chi = n^{-1} \sum_i |1 - y_i/x_i|$ between the measured inductances, $x_i$, and the calculated values, $y_i$; $n$ is the number of inductors in the set. A set of composite inductors M6aM4 located on two adjacent 5 mm x 5 mm test chips was used. Geometrical parameters of these inductors are given in Table II; they all have $N_{sq} = 10$ squares of NbN in the middle of the inductors.

TABLE II
Parameters of composite NbN+NbN/Nb bilayer inductors M6aM4

| No | $w$ (μm) | $l$ (μm) | $l_{NbN}$ (μm) | $L_{meas}$ [a] (pH) | $L_{meas}$ [b] (pH) | $L_{calc}$ [c] (pH) |
|---|---|---|---|---|---|---|
| 1 | 0.25 | 15 | 2.5 | 36.03 | 35.61 | 34.36 |
| 2 | 0.35 | 20 | 3.5 | 36.64 | 36.26 | 36.13 |
| 3 | 0.50 | 20 | 5.0 | 39.03 | 31.26 | 33.34 |
| 4 | 0.70 | 25 | 7.0 | 33.66 | 32.53 | 34.01 |
| 5 | 1.00 | 30 | 10.0 | 31.75 | 31.76 | 33.81 |
| 6 | 2.00 | 55 | 20.0 | 35.17 | 36.75 | 36.75 |
| 7 | 4.00 | 95 | 40.0 | 36.14 | 38.87 | 37.79 |

[a] Measured inductance of M6aM4 composite microstrips on wafer 6 at the chips location D4 (w6D4); the NbN/Nb bilayer has $t_{NbN} = 150$ nm and $t_{Nb} = 100$ nm.
[b] Measured inductance of M6aM4 composite microstrips on w6E4
[c] Calculated using the kinetic inductance of the NbN layer of $L_{K,sq} = 2.055$ pH/sq, giving MAD values $\chi = 5.3\%$ and 3.5% for the w6D4 and w6E4 data, respectively. The best fit for w6D4 data gives $L_{K,sq} = 2.02$ pH/sq.

TABLE III
PARAMETERS OF COMPOSITE NbN+NbN/Nb BILAYER MICROSTRIP M6bM7
AND STRIPLINE M6aM4bM7 INDUCTORS

| No | $l_{NbN}$[a] (μm) | $L_{meas}$[b] (pH) | $L_{meas}$[c] (pH) | $l_{NbN}$[d] (μm) | $L_{meas}$[e] (pH) | $L_{meas}$[f] (pH) |
|---|---|---|---|---|---|---|
| 1 | 1.5 | 31.40 | 28.56 | 1.0 | 27.42 | 26.02 |
| 2 | 2.1 | 30.20 | 29.79 | 1.4 | 26.84 | 24.83 |
| 3 | 3.0 | 26.60 | 24.10 | 2.0 | 21.17 | 20.41 |
| 4 | 4.2 | 24.16 | 23.77 | 2.8 | – | 19.98 |
| 5 | 6.0 | 24.13 | 22.59 | 4.0 | 20.05 | 18.84 |
| 6 | 12.0 | 23.44 | – | 8.0 | 19.86 | 18.47 |
| 7 | 24.0 | 22.24 | 23.13 | 16.0 | 17.79 | 18.24 |

[a] Lengths of the NbN part of the M6bM7 microstrips. The total lengths and the linewidths of the inductors are given in Table II on the same numbered rows.
[b] Measured inductance of the M6bM7 composite microstrips at location D4 on wafer #6, abbreviated w6D4.
[c] The measured inductance of the M6bM7 composite microstrips on w6E4.
[d] Lengths of the NbN part of the M6aM4bM7 striplines. The total lengths and the linewidths of the inductors are given in Table II on the same numbered rows.
[e] Measured inductance of the M6aM4bM7 composite inductors on w6D4.
[f] Measured inductance of the M6aM4bM7 composite inductors on w6E4.

Inductances of the NbN and Nb layers of the bilayers, required for (7), and the geometrical inductance of the NbN $L_{G,l}^{NbN}$ were calculated using (1) for the microstrips and (2) or (3) for the striplines, with the appropriate thicknesses and linewidths. These inductances were also extracted using wxLC [24], giving the same results. The mutual inductance $M_{NbN\text{-}Nb}$ was calculated using (4) and (5), and the more general expressions in [6], and also using wxLC, or InductEx [25]. It is important to note that $M_{NbN\text{-}Nb}$ does not depend on $\lambda_{NbN}$ and $\lambda_{Nb}$; it depends only on the geometrical dimensions and the penetration depth in the ground planes $\lambda$ [6].

Since $\lambda_{NbN} \gg \lambda_{Nb}$, $L_{NbN} \gg L_{Nb}$ whereas $M_{NbN,Nb}$ is the smallest of all the inductances. Hence, inductance of the bilayer is approximately equal to the inductance of the top Nb layer, $L_{l,bi} \approx L_{l,Nb}(1 - L_{l,Nb}/L_{l,NbN} + 2M_{l,NbN\text{-}Nb}/L_{l,NbN})$ and practically insensitive to the exact value of $\lambda_{NbN}$ as long as $\lambda_{NbN} \gg \lambda_{Nb}$. Geometrical inductance $L_{G,l}^{NbN}$ also does not depend on $\lambda_{NbN}$. Therefore, the only important adjustable parameter in (7) is kinetic inductance $L_{K,sq}^{NbN}$.

The obtained values of $\lambda_{NbN}$ and $L_{K,sq}^{NbN}$ were used to calculate the sum of the bilayer inductance and the geometrical inductance of the NbN parts, $L_{bi} + L_G^{NbN}$, for three other sets of $n=7$ inductors measured on the same chips: M6bM7 microstrips with the same widths and lengths as in Table II but with 6 squares of NbN; M6aM4bM7 striplines with the same widths and lengths as in Table II but with $N_{sq} = 4$, and M6aM4bM7 2-μm-wide striplines (M6aM4bM7_2μm) with $N_{sq} = 1$.

The measured inductances for the M6bM7 composite microstrips and M6aM4bM7 striplines, along with the corresponding lengths of the NbN part, are shown in Table III for the location D4 and E4 on wafer #6 (w6). The measurements were repeated on the same sets of inductors at four locations, spaced 22 mm apart in the radial direction starting from the wafer

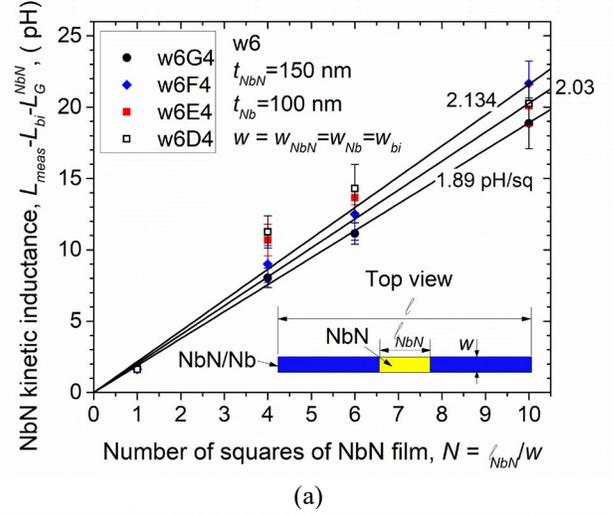

(a)

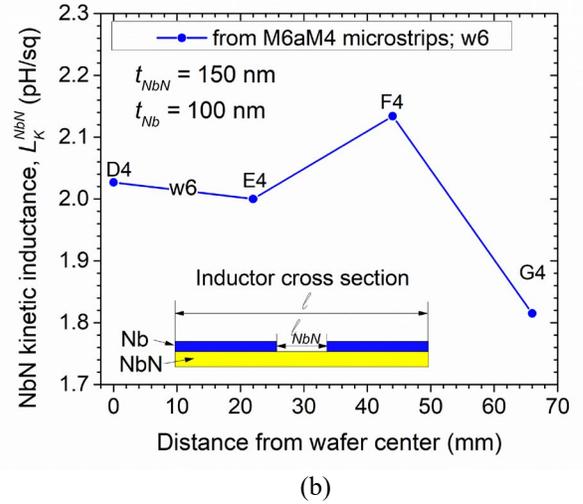

(b)

Fig. 8. (a) Kinetic inductance of the NbN part of composite microstrip inductors M6aM4 ($N$=10) and M6bM7 ($N$=6), and composite striplines M6aM4bM7 ($N$=4) at four locations on w6; (b) wafermap of the kinetic inductance. NbN kinetic conductance was calculated by deducting the calculated inductance of the bilayer parts, $L_{bi}$ and the geometrical inductance of the NbN part, $L_G^{NbN}$ in (8) from the measured total inductance $L_{meas}$; see (9). The slope of the linear dependences in (a) gives kinetic inductance per square, $L_{K,sq}^{NbN}$ at four different locations on the wafer. Insets show the composite inductor top view (a) and the corss section (b).

center (location D4) to the right wafer edge (location G4), on two wafers, w5 and w6.

NbN/Nb bilayer on w6 was deposited with thicknesses $t_{NbN} = 150$ nm and $t_{Nb} = 100$ nm. For calculating inductance of the bilayer microstrips M6aM4, using (1)–(5), $d_1 = 615$ nm for the NbN layer and $d_2 = d_1 + t_{NbN} = 765$ nm for the Nb layer. For the inverted bilayer microstrips M6bM7 with the M7 ground plane, $d_1 = 200$ nm for the Nb layer and $d_2 = 200 + t_{Nb} = 300$ nm for the NbN layer. Inductances of the composite microstrips M6aM4, calculated using (7), are given in Table II at the $L_{K,sq}^{NbN} = 2.055$ pH/sq, corresponding to $\lambda_{NbN} = 495$ nm.

The extracted kinetic inductance of the NbN part of the measured composite inductors is shown in Fig. 8a for M6bM7

microstrips with $N_{sq}$=6, M6aM4bM7 striplines with $N_{sq}$=4 and $N_{sq}$=1. It was extracted as

$$L_K^{NbN} = L_{meas} - L_{bi} + L_G^{NbN}, \quad (9)$$

where the inductance of the bilayer parts $L_{bi}$ and geometrical inductance of the NbN part $L_G^{NbN}$ was calculated as described above, using $\lambda_{NbN}$ determined from the fits to the data on M6aM4 microstrips with $N_{sq}$=10. The corresponding kinetic inductance per square $L_{K,sq}^{NbN}$ is given in Fig. 8a by the slope of the straight lines shown in Fig. 8a.

Fig. 8b shows that there is some variation of the kinetic inductance across the wafers, which will be discussed in Sec. V. The mean value for two wafers measured is $L_{K,sq}^{NbN}$ =2.08 pH/sq, which is consistent with the expected value of 2.01 pH/sq based on the measurements in Sec. III of the single layer 200-nm-thick NbN inductors with Nb ground plane(s).

The described measurements were repeated on wafers with different NbN/Nb bilayer thicknesses: 100 / 100 nm and 130 / 70 nm. We do not present these data and the extraction results for the lack of space.

## V. DISCUSSION OF NbN KINETIC INDUCTANCE

There have been numerous measurements of the penetration depth in various NbN films, showing values from about 100 nm in the epitaxial single-crystal films up to 600 nm in polycrystalline films [26]–[33], depending on the resistivity, deposition method, substrate material, deposition temperature, and microstructure of the films. Very thin, single-layer NbN films have widely been used for photon counters – superconducting nanowire single photon detectors [34]–[37] and microwave kinetic inductance detectors; see [38] and references therein. The large variation in $\lambda_{NbN}$ values in NbN films is attributed to a large variation in the films disorder, effects of localization and formation of a nonuniform superconducting state; see [39] for the most recent review.

In the microscopic theory of superconductivity [40], the penetration depth in weak coupling superconductors with short mean free path of electrons is given by [41], [42]

$$\lambda_{BCS}(T) = \left(\frac{\hbar}{\mu_0 \pi \sigma \Delta \tanh\frac{\Delta}{2k_B T}}\right)^{1/2}, \quad (10)$$

where $\Delta$ is the superconducting energy gap, $\sigma$ the normal state conductivity, $\hbar$ the reduced Planck's constant, and $k_B$ Boltzmann constant. In this model, the kinetic inductance of a superconducting film per square is

$$L_K^{BCS} = \mu_0 \frac{\lambda_{BCS}^2}{d} = R_{sh} \frac{\hbar}{\pi \Delta(T) \tanh\frac{\Delta(T)}{2k_B T}}, \quad (11)$$

where $R_{sh} \equiv 1/(\sigma d)$ is the sheet resistance of the film in the normal state. For NbN films at 4.2 K, $T \ll T_c$, $\Delta(T) \approx \Delta(0) = 1.764\eta_\Delta k_B T_c$, and (11) reduces to

$$L_K^{BCS} \text{ [pH/sq]} = 1.3783 R_{sh}/(\eta_\Delta T_c), \quad T \ll T_c, \quad (12)$$

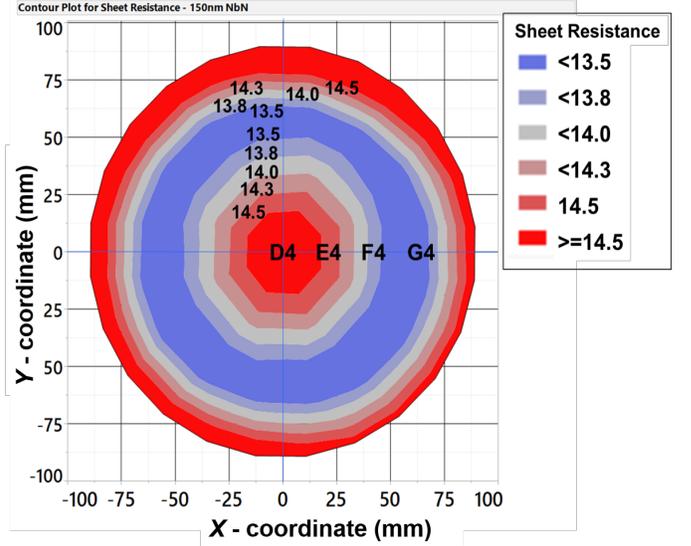

Fig. 9. Sheet resistance $R_{sh}$(300 K) wafermap of a 150 nm NbN film on a 200-mm oxidized Si wafer. Low temperature sheet resistance is $R_{sh}$(20 K)= $R_{sh}$(300 K)/$r$ and has the same distribution on the wafer.

where $R_{sh}$ is in ohms and $T_c$ in kelvin; $\eta_\Delta = \Delta(0)/1.764 k_B T_c$ characterizes deviation of the $\Delta(0)/k_B T_c$ of the superconductor from the BCS theory value 1.764 [40]. So, the kinetic inductance of a superconducting film is directly proportional to the film sheet resistance and should have the same distribution across the wafer if $T_c$ of the film, mainly governed by the film composition, is relatively constant.

NbN is viewed a strong coupling superconductor with $\Delta/k_B T_c = 2.155$ [43], i.e., $\eta_\Delta$ =1.222. Therefore, for uniform NbN films at $T \ll T_c$, we may expect that

$$L_{K,sq}^{NbN} \text{ [pH/sq]} = \frac{1.1279 R_{sh} \text{ [}\Omega\text{/sq]}}{T_c \text{ [K]}}. \quad (12)$$

The wafermap of the sheet resistance of a 150-nm-thick NbN film at room temperature, $R_{sh}$(300 K) is shown in Fig. 9. It has a nearly circular symmetry and the radial change of the sheet resistance resembling the erosion pattern of the sputtering target of the Nb deposition magnetron. The ratio of the resistances at the D4 and G4 locations is about 1.11. Similarly, the ratio of the NbN kinetic inductances on these two sites on w6 is 1.13, close to the ratio of the film's sheet resistances.

Using the measured $R_{sh}$ (300 K) from the wafermap in Fig. 9, the measured resistance ratio $r = R(300K)/R(20K)$, and $T_c$ values given in Table IV, the BCS theory-based (11) predicts noticeably lower kinetic inductance values, $L_K^{BCS}$, than the measured values; see Table IV. Using (12) would give a factor $\eta_\Delta$ yet lower values. The difference indicates that the superfluid density in the disordered NbN films, $n_s$ is noticeably lower than the expected density in the conventional disordered superconductors $n_s^{BSC} = \pi \sigma \Delta m^*/(\hbar e^2)$ [41], where $m^*$ and $e$ are the effective mass and electron charge, respectively. The ratio $n_s/n_s^{BSC} = L_K^{BCS}/L_K^{meas}$ is shown in Table IV. Interestingly, we observed even a larger reduction in the superfluid density in thin MoN$_x$ films [11, Fig. 8]. The 40-nm-thick MoN$_x$ films with $x \approx 0.5$, used in the SFQ5ee process [9] as kinetic inductors, have $R_{sh}$(300 K)=25 $\Omega$/sq, $T_c$ =7.5 K and $L_{K,sq}^{Mo2N}$ =8.0 pH/sq,

TABLE IV
ELECTRIC PARAMETERS AND KINETIC INDUCTANCE OF NbN FILMS

| Location | $R_{sh}$ (Ω) 300 K | $R(300)/R(20)$ | $R_{sh}$ (Ω) at 20 K | $T_c$ (K) | $L_K^{BCS}$ (pH/sq) | $\frac{n_s}{n_s^{BCS}}$ [c] |
|---|---|---|---|---|---|---|
| D4[a] | 15.0 | 0.8589 | 17.46 | 15.4 | 1.605 | 0.752 |
| E4[a] | 14.5 | 0.8548 | 16.98 | 15.3 | 1.528 | 0.764 |
| F4[a] | 14.0 | 0.8507 | 16.46 | 15.2 | 1.492 | 0.735 |
| G4[a] | 13.5 | 0.8466 | 19.95 | 15.1 | 1.455 | 0.770 |
| F4[b] | 8.4 | 0.8507 | 9.88 | 15.2 | 0.884 | 0.593 |

[a] For 150-nm-thick NbN film
[b] For 200-nm-thick NbN film
[c] The ratio calculated using the BCS theory-based (11) value and the measured kinetic inductance

whereas (11) predicts the sheet kinetic inductance of only 4.6 pH/sq, giving $n_s/n_s^{BSC} = 0.575$. The ratio decreases with decreasing the nitrogen content in the MoN$_x$, much faster than $T_c$ decreases [11, Fig. 8].

Reduction of the superfluid density in disordered superconductor is caused by Anderson localization of the current carriers which competes with Cooper paring [44]. More recently, this problem was studied in the framework of the bosonic mechanism of the superconductor-insulator transition; see [45], [46], [47], and references therein. It was shown that, with increasing disorder, the superfluid density decreases proportionally to $\Delta^2 R_0^2$, instead of proportionally to $\Delta$ as in the BCS theory, where $R_0$ is the spatial range of tunneling matrix elements between the localized states. Experimental data on tunneling in disordered NbN films can be found in [48], [49]. The only evidence that our films are in the weak localization regime is that their resistance slightly increases with decreasing temperature, $r < 1$. However, high $T_c$ values and a relatively small reduction in $n_s$ with respect to the $n_s^{BCS}$, see Table IV, indicate that the films are far from the superconductor-insulator transition. It remains to be studied if disorder and localization effects may affect reproducibility of the kinetic inductance with decreasing the inductors dimensions, and thereby impede the use of disordered kinetic inductors in very large-scale integrated circuits.

## VI. CONCLUSION

The main goal of this work was to show that kinetic inductors can be implemented in a multilayered fabrication process and used as cell inductors in integrated circuits. We presented kinetic inductance and mutual inductance data for NbN microstrips and striplines with $T_c \approx 16$ K incorporated in a fully planarized multilayered fabrication process with Nb ground planes and Nb/Al-AlO$_x$/Nb Josephson junctions, an advanced node of the standard SFQ5ee process. We also presented inductance data for various composite inductors Nb + NbN/Nb developed for compact inductors in logic cells of digital and mixed-signal integrated circuits. We also measured inductance of 90-degree bends and found that it can be neglected for thin-film, $t < \lambda_1$, kinetic (NbN) inductors with $w \ll \lambda_1^2/t$ and geometrical (Nb) inductors with $w < d_1$ in the presence of superconducting ground plane(s), i.e., the corner inductance to the sheet inductance ratio $\alpha = L_c/L_{sq} \ll 1$ is much smaller than that expected from the current crowding near the inner corners of the single-layer structures with the bends [22], [23], [58], [59]. This result agrees with the previous measurements in [6] on the Nb meandered striplines.

Our measurements indicate good reproducibility of the NbN and of composite NbN+NbN/Nb inductors from wafer to wafer, and provide inductance data for calibration of the inductance extractors such as the newest InductEx 7.0 [25]. The presented data should be sufficient for the circuit design in the advanced nodes of superconductor electronics processes utilizing NbN inductors, developed at MIT LL.

Another material for incorporation in our process and integration with Nb/Al-AlO$_x$/Nb junctions could be NbTiN films [50]−[53], which have a bit smaller penetration depth than NbN, in the range from 230 nm to 360 nm, depending on the deposition conditions [51], [53], and could be more convenient in this respect and from the point of view of NbTiN materials properties. On the other hand, the use in logic cells of kinetic inductors with larger sheet inductances than about 3 pH/sq, e.g., 40-nm-thick Mo$_2$N inductors with $L_{K,sq} = 8$ pH/sq developed for the SFQ5ee process, is not convenient because high values of $L_{K,sq}$ result in too short traces of inductors which are difficult to pattern and control reliably. Also, very thin kinetic inductors have low critical currents, inversely proportional to their inductance, especially in vias, which limit their usefulness for integrated circuits.

One of the main beneficiaries of kinetic inductors will be energy-efficient integrated circuits using Josephson junction with low critical currents, below about 50 μA, which require large inductance values of their cell inductors. For instance, superconductor neuromorphic circuits have recently become a subject of hot interest; see a review [54] and references therein. Many types of neuromorphic circuits process information in a stochastic manner and are much less sensitive to bit error rates than digital circuits; see [55]–[57]. Hence, critical currents of the JJs in them, and the energy dissipation, can be substantially lowered, down to 10−20 μA, with respect to the typical critical currents $I_c \sim 100$ μA used in digital circuits. This necessitates a proportional, a 5× to 10×, increase of the cell inductances, and can only be realized using kinetic inductors.

Full implementation of NbN/Nb bilayer inductors should provide for a 10× increase in the inductor number density and thereby significantly increase integration scale of superconductor electronics. Further increase in the circuit density would require reducing the size of Josephson junctions and interlayer vias; see the density estimates and requirements in [1] and [60].

## ACKNOWLEDGMENT

We are grateful to Vasili Semenov for numerous discussions of bilayer inductors and neuromorphic circuits, and to Ravi Rastogi for his help in wafer processing. S.K. Tolpygo would like to thank Mikhail M. Khapaev for the access to inductance extraction software wxLC, and to Coenrad J. Fourie for the access to and help with InductEx. We would like to thank Leonard Johnson and Mark Gouker for their interest in this research.

This research was based upon work supported by the Under Secretary of Defense for Research and Engineering via Air Force Contract No. FA8702-15-D-0001. Any opinions, findings, conclusions or recommendations expressed in this material are those of the authors and do not necessarily reflect the views of the Under Secretary of Defense for Research and Engineering and should not be interpreted as necessarily representing the official policies or endorsements, either expressed or implied, of the U.S. Government. Delivered to the U.S. Government with Unlimited Rights, as defined in DFARS Part 252.227-7013 or 7014 (Feb 2014). Notwithstanding any copyright notice, U.S. Government rights in this work are defined by DFARS 252.227-7013 or DFARS 252.227-7014 as detailed above. Use of this work other than as specifically authorized by the U.S. Government may violate any copyrights that exist in this work. The U.S. Government is authorized to reproduce and distribute reprints for Governmental purposes notwithstanding any copyright annotation thereon.